# Analysis of the $\pi^0\pi^0$ Final State in $\pi^-p$ Reactions at $18.3\,GeV/c$

J. Gunter[1], G. S. Adams[6], T. Adams[2], E. V. Anoshina[8], Z. Bar-Yam[4], J. M. Bishop[2], V. A. Bodyagin[8], B. B. Brabson[1], D. S. Brown[5], N. M. Cason[2], S. U. Chung[3], R. R. Crittenden[1], J. P. Cummings[4], S. Denisov[7], J. P. Dowd[4], A. Dushkin[7], A. R. Dzierba[1], P. Eugenio[4], A. M. Gribushin[8], R. W. Hackenburg[3], M. Hayek[4], W. Kern[4], E. King[4], V. Kochetkov[7], O. L. Kodolova[8], V. L. Korotkikh[8], M. A. Kostin[8], R. Lindenbusch[1], J. M. LoSecco[2], J. J. Manak[2], J. Napolitano[6], M. Nozar[6], C. Olchanski[3], A. I. Ostrovidov[8], T. K. Pedlar[5], A. S. Proskuryakov[8], D. R. Rust[1], A. H. Sanjari[2], L. I. Sarycheva[8], E. Scott[1], K. K. Seth[5], I. Shein[7], W. D. Shephard[2], N. B. Sinev[8], J. A. Smith[6], P. T. Smith[1], A. Soldatov[7], D. L. Stienike[2], T. Sulanke[1], S. A. Taegar[2], S. Teige[1], D. R. Thompson[2], I. N. Vardanyan[8], D. P. Weygand[3], H. J. Willutzki[3], J. Wise[5], M. Witkowski[6] A. A. Yershov[8], D. Zhao[5],

[1] Department of Physics, Indiana University,Bloomington IN 47405, USA
[2] Department of Physics, University of Notre Dame, Notre Dame IN 46556, USA
[3] Department of Physics, Brookhaven National Laboratory,Upton, L.I., NY 11973
[4] Department of Physics, University of Massachusetts Dartmouth,North Dartmouth, MA 02747,USA
[5] Department of Physics, Northwestern University,Evanston, IL 60208, USA
[6] Department of Physics, Ressselaer Polytechnic Institute, Troy NY,USA
[7] Institute for High Energy Physics,Protovino, Russian Federation
[8] Institute for Nuclear Physics, Noscow State University,Moscow, Russian Federation

The reaction $\pi^-p \to \pi^0\pi^0 n$ may be used to examine both scalar and tensor mesons. The $f_0(980)$ signature changes from destructive interference at small momentum transfer $(-t < 0.1\,GeV^2)$ to an enhancement at larger momentum transfer. At small momentum transfer one pion exchange allows extraction of $\pi\pi$ phase shifts and inelasticities. The $f_2(1270)$ production mechanism also changes as a function of momentum transfer. Unnatural parity exchange dominates $f_2(1270)$ production at at small momentum transfer while natural parity exchange becomes the leading production mechanism at larger momentum transfer. These results are based on an analysis of 188,000 $\pi^-p \to \pi^0\pi^0 n$ events collected by experiment E852 at Brookhaven National Laboratory during the 1994 HEP running period.

## 1  Data Sample

The $\pi^-p \to \pi^0\pi^0 n$ signal is extracted from events with four photons and no charged tracks in the 1994 E852 data set. A sample of 188,000 events are obtained. Photons are detected by a large lead glass calorimeter [1,2]. Events are selected via a three constraint kinematic fit. The three constraint $\chi^2$ for the $\pi^0\pi^0 n$ topology is required to be less than 7.8 (95% C.L.) while no other topology (such as $\eta\pi^0 n$ or $\eta\eta n$) has a smaller $\chi^2$. To enhance the exclusivity of the data set a requirement is made that the Cesium Iodide [3] veto barrel surrounding the target detect less than $20\,MeV$ visible energy.

## 2  Description of Mass and Momentum Transfer Distributions

The momentum transfer dependence is characterized by a steeply falling region $(\exp(-8|t|))$ for $|t| < 0.2\,GeV^2$ and shallower $(\exp(-4|t|))$ region at larger values of $|t|$. The mass distribution of events with $0.03 < |t| < 0.10\,GeV^2$ is dominated by the $f_2(1270)$ and has a sharp dip at the $f_0(980)$. The region of $0.40 < |t| < 1.50\,GeV^2$ is dramatically different. The $f_2(1270)$ signal persists and a narrow bump appears at $980\,MeV$ as has been observed by the GAMS collaboration [4].

## 3  Partial Wave Analysis

Partial wave analyses have been performed on this system for various $|t|$ ranges. Within each $|t|$ range data are binned in $20\,MeV$ mass bins. The partial wave analysis consists of using an extended maximum likelihood method fit to decompose the observed angular distribution in the Gottfried-Jackson frame for each (m,t) bin into partial waves. Partial waves with a subscript "0" or "-" correspond to unnatural parity exchange processes, while a "+" subscript indicates natural parity exchange. The partial wave decomposition is not unique, that is, ambiguities exist. If only $S_0, D_0, D_+,$ and $D_-$ waves are included in the fit, there are at most two ambiguous solutions.

The results of the partial wave decomposition for events in the range $0.40 < |t| < 1.50\,GeV^2$ are shown in figure 3.1. The $f_2(1270)$ meson is observed in both the $D_+$ and $D_0$ waves, with the peak $D_+$ intensity approximately twice as large as the peak intensity in the $D_0$ partial wave. Thus, both natural and unnatural parity exchange are important. The narrow bump which is observed in the mass spectrum near $1.0\,GeV$ is observed in the $S_0$ wave which also shows a broad enhancement peaking near 1350 MeV. The ambiguous set of solutions is quite similar and not shown.

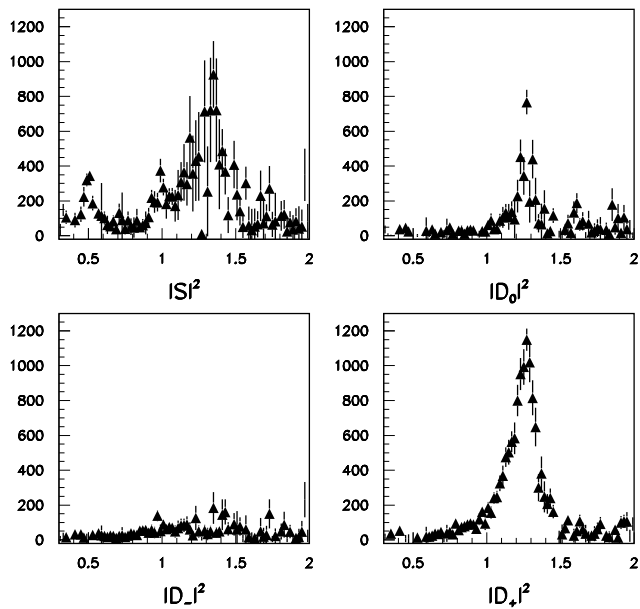

**Figure 3.1** Partial wave analysis results for $0.40 < |t| < 1.50\,GeV^2$ in 20 Mev $\pi\pi$ mass bins.

Figure 3.2 shows the results of partial wave decompostion of events with $0.03 < |t| < 0.10\,GeV^2$. Both ambiguous solutions are shown. Here the $f_2(1270)$ is dominantly found in the $D_0$ partial wave, consistent with the dominance of a one pion exchange (OPE) mechanism in this region. Historically, no large spin-2 intensity has been found under the $\rho^0$ meson in analyses of the $\pi^+\pi^-$ final state.[5] Thus, in this region, the solution with larger $D_0$ intensity is disfavored. No similar argument may be applied to the region above $K\bar{K}$ threshold. The S-wave intensity associated with the favored $D_0$ intensity below this threshold shows very broad structure leading into a narrow dip. In both solutions the $S_0$ intensity rises again, peaking near $1.3\,GeV$.

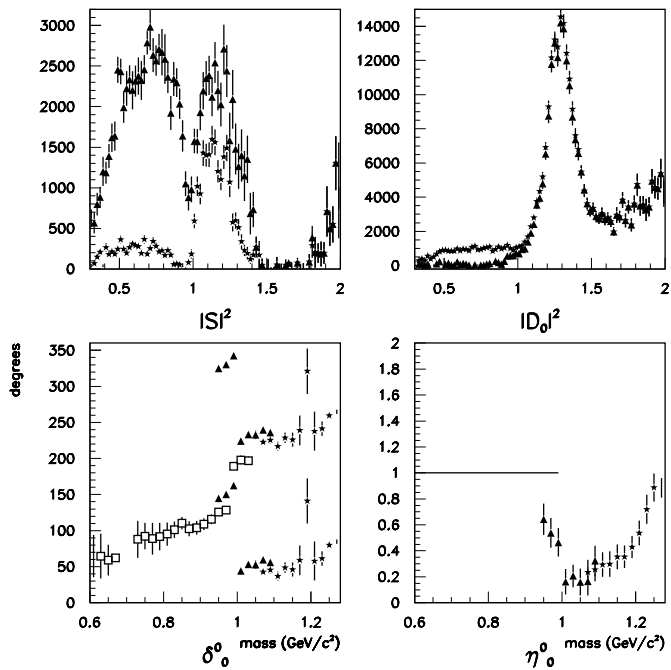

**Figure 3.2** Results for $0.03 < |t| < 0.10\,GeV^2$ in 20 Mev $\pi\pi$ mass bins. Square symbols in the lower lefthand plot are determined from the $|S|^2$ alone under the assumption $\eta_0^0$ is unity.

OPE dominance at small momentum transfer allows the reaction $\pi^- p \to \pi^0 \pi^0 n$ to be thought of as $\pi\pi$ scattering [6]. Interest in $I = J = 0$ mesons exists as this sector is a potential hunting ground for non-$q\bar{q}$ mesons. The physical $\pi^0 \pi^0$ state, however, contains both I=0 and I=2 components. If the I=2 behavior is taken from existing measurements of $\pi^+ p \to \pi^+ \pi^+ n$ [7] the I=0 behavior may be isolated. The interesting parameters to be determined for I=J=0 $\pi\pi$ scattering are the phase shift $\delta_0^0$ and inelasticity $\eta_0^0$.

Below $K\bar{K}$ threshold the inelasiticity $\eta_0^0$ may be assumed to be unity. This leaves only $\delta_0^0$ to be found in this region. The $S_0$ wave intensity $|S_0|^2$ is sufficient to determine $\delta_0^0$. Above $K\bar{K}$ threshold $\eta_0^0$ is allowed to vary, increasing the amount of necessary information. Although the PWA cannot determine an absolute phase, by using the relative phase of the $S_0$ and $D_0$ waves and assum-ing the $D_0$ wave is dominated by the $f_2(1270)$ resonance, the $S_0$ wave phase $\phi_S$ may be deduced. Thus, $\delta_0^0$ (modulo $180°$) and $\eta_0^0$ maybe determined from $|S|^2$ and $\phi_S$. The resulting phase shifts and inelasticities for $0.6 < M_{\pi\pi} < 1.3\,GeV$ are shown in figure 3.2. Above $1.3\,GeV$ the low mass tail of the $f_4(2040)$ causes the $D_0$ wave (and hence the calculated $\phi_S$) to become uncertain. The phase shift $\delta_0^0$ climbs rapidly through $180°$ near $980\,MeV$, consistent with the presence of the $f_0(980)$, while $\eta_0^0$ indicates the strong opening of the $K\bar{K}$ exit channel.

## References

1. B. Brabson *et al.*, *Nucl. Instrum. Methods* **A 332**, 419 (1993).
2. R. R. Crittenden *et al.*, submitted to *Nucl. Instrum. Methods*.
3. T. Adams *et al.*, *Nucl. Instrum. Methods* **A 368**, 617 (1996).
4. D. Alde it et al., *Z. Phys.* **C 66**, 375 (1995).
5. G. Grayer *et al.*, *Nucl. Phys.* **B 75**, 189 (1974).
6. P. Estabrooks and A. D. Martin, *Nucl. Phys.* **B 95**, 322 (1975).
7. W. Hoogland *et al.*, *Nucl. Phys.* **B 126**, 109 (1977).